\documentclass[reprint,superscriptaddress]{revtex4-1}

\usepackage{graphicx}
\usepackage{amsmath}

\def\voc{\mathrel{\rlap{\lower0pt\hbox{\hskip1pt{$c$}}}\raise3pt\hbox{$\neg$}}}

\begin{document}

\title{Bounds on higher-order Lorentz-violating photon sector coefficients from an asymmetric optical ring resonator experiment}

\author{Stephen R. Parker}
\email{stephen.parker@uwa.edu.au}
\affiliation{School of Physics, The University of Western Australia, Crawley 6009, Australia}
\author{Matthew Mewes}
\affiliation{California Polytechnic State University, San Luis Obispo, California 93407, USA}
\author{Fred N. Baynes}
\author{Michael E. Tobar}
\affiliation{School of Physics, The University of Western Australia, Crawley 6009, Australia}

\date{\today}

\begin{abstract}
Optical resonators provide a powerful tool for testing aspects of Lorentz invariance. Here, we present a reanalysis of an experiment where a path asymmetry was created in an optical ring resonator by introducing a dielectric prism in one arm. The frequency difference of the two fundamental counter-propagating modes was then recorded as the apparatus was orientation-modulated in the laboratory. By assuming that the minimal Standard-Model Extension coefficients vanish we are able to place bounds on higher-order parity-odd Lorentz-violating coefficients of the Standard-Model Extension. The results presented in this work set the first constraints on two previously unbounded linear combinations of $d=8$ parity-odd nonbirefringent nondispersive coefficients of the photon sector.
\end{abstract}



\maketitle

\section{Introduction}
Lorentz invariance is a fundamental component of the Standard Model and General Relativity. Despite the successes of both theories they remain incompatible; it is generally assumed they are both low-energy approximations of a single theory that is consistent at the Planck scale~\cite{tasson2014}. Various efforts towards identifying a unified theory can allow or require Lorentz invariance to be broken~\cite{tasson2014,Jacobson2006}. Testing Lorentz invariance thus provides one of the few experimental portals for assessing and guiding theories of quantum gravity and other unification propositions. Precision laboratory measurements of Planck-scale suppressed effects offer excellent prospects in the search for Lorentz-Invariance Violations (LIV)~\cite{tasson2014}.
\\
The Standard-Model Extension (SME) provides a comprehensive framework for analysing, quantifying and comparing different experimental tests of LIV~\cite{ck1,ck2,Kostelecky:2002,datatables}. The SME describes all possible Lorentz and $CPT$ violations associated with known particles and fields. Efforts have usually focused on the minimal SME sectors, which only contain energy-independent operators of renormalizable dimension in flat spacetime. In recent years the SME has been expanded to include higher-order nonrenormalizable operators, which presents new opportunities for experimentation and analysis~\cite{kostelecky2009,mewes1}. Bounds have previously been placed in the photon sector~\cite{steve1,yuta2}, with results in other sectors now starting to emerge~\cite{HOmuons,HOneutrinos}. Practically there are situations where one might expect LIV to manifest at higher-order, e.g. in some theories with noncommutative spacetime coordinates LIV only occurs for nonrenormalizable operators~\cite{Hayakawa2000,carroll2001}.
\\
In the photon sector of the SME operators can be classified into different groups that describe their effects on standard electrodynamics. Astrophysical observations have tightly constrained birefringent and vacuum dispersive effects with sensitivities far beyond the reach of terrestrial tests. This provides physical motivation to the study of the camouflage coefficients, which to leading order are nonbirefringent and nondispersive. The camouflage coefficients are $CPT$ invariant, have even dimension $d\geq4$, and are best constrained via precision electromagnetic resonant-cavity tests, such as the modern descendants of the Michelson-Morley experiment.
\\
Constraints have previously been set on combinations of $d=6$ and $d=8$ camouflage coefficients using data from a parity-even microwave cavity experiment~\cite{steve1} and a parity-odd optical cavity experiment~\cite{yuta2}. As the sensitivity to the camouflage coefficients scales with the frequency of the photon (roughly as $\nu^{d-4}$), optical cavities are very well suited for testing these effects~\cite{mewes1}. For the parity-odd camouflage coefficients there are three $d=6$ coefficients and 13 $d=8$ coefficients, with three linear combinations currently unbounded. Here we analyse an orientation modulated parity-odd asymmetric optical ring resonator experiment to set bounds on higher-order parity-odd camouflage coefficients, including the first constraints on two of the three remaining unbounded $d=8$ combinations.
\section{Experimental Setup}
\begin{figure*}[t!h]
\centering
\includegraphics[width=1.5\columnwidth]{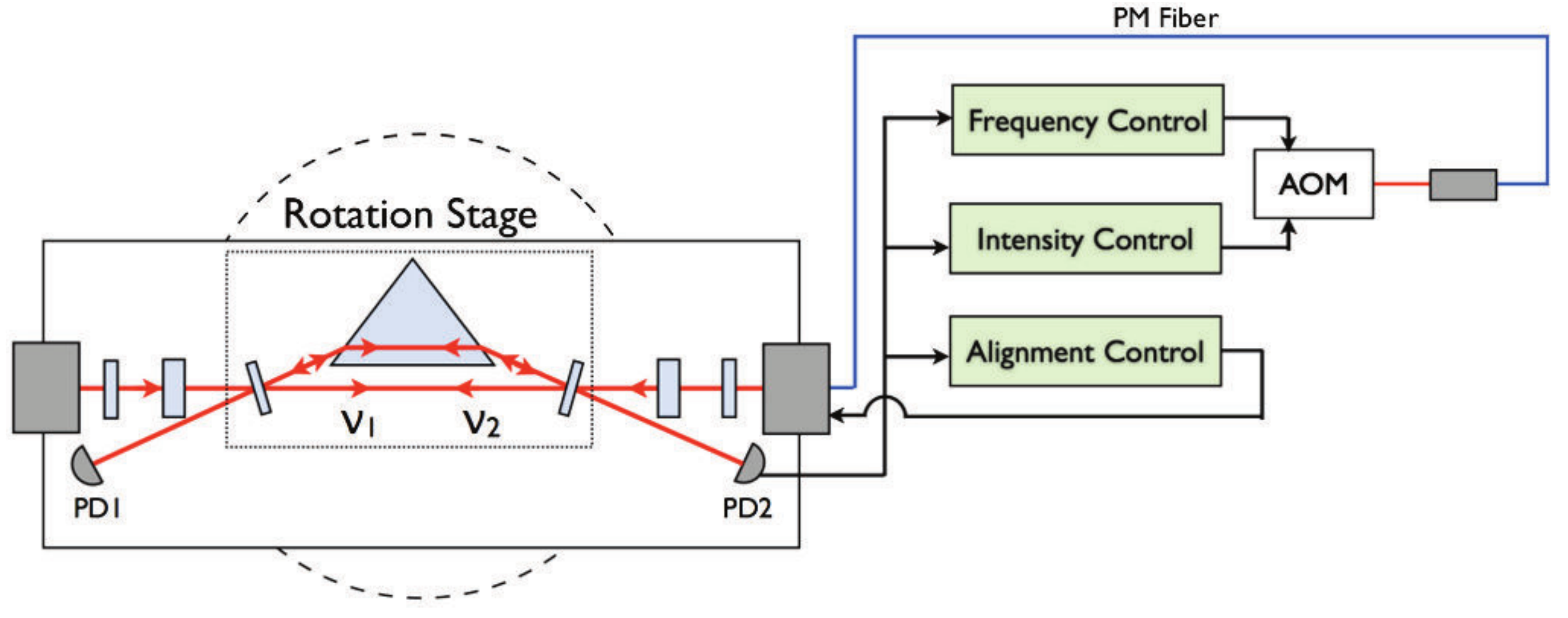}
\caption{Schematic of the parity-odd asymmetric optical ring resonator experiment, reproduced from~\cite{fred2}. This only shows the control systems and setup for one of the counter-propagating modes, there is a nominally identical setup for the second mode.}
\label{fig:exp}
\end{figure*}
A detailed overview of the experiment is provided in~\cite{fred1} and~\cite{fred2}. As seen in Fig.~\ref{fig:exp} a dielectric prism with index of refraction $n=1.44$ is placed in one arm of an optical ring resonator to create a path asymmetry. A 1064~nm laser is split and each path is independently locked to a fundamental counter-propagating mode. The first laser path is frequency shifted by an Acousto-Optic Modulator (AOM), introducing a 160~MHz offset, and then mode-locked by applying correction signals to the laser. The second path is locked to the counter-propagating mode via an AOM; the correction signal contains the frequency difference (typically $\sim$100~Hz) of the two modes imprinted on the 160~MHz offset. This was recorded directly with a frequency counter.
\\
The apparatus sits upon a rotation platform. The beat frequency was recorded for 10 minutes in a stationary position, then the experiment was rotated by +180$^{\circ}$ and the beat frequency recorded for 10 minutes while stationary. The experiment was then rotated back by -180$^{\circ}$ to the initial position and the cycle repeated. The result of this is that the final dataset is in essence stationary, but with leading-order noise processes shifted to a different frequency. Previously, measurements were made for 50 days and used to place bounds in the minimal SME~\cite{fred2}. We use the same dataset for the following analysis, but assume that the minimal SME terms vanish.
\\
\section{Experimental Sensitivity to Higher-Order Camouflage Coefficients}

A complete derivation of the camouflage coefficients is provided in~\cite{kostelecky2009}. In short, the coefficients associated with $CPT$-even differential operators from the arbitrary Lorentz and $CPT$ violating Lagrange density for photon propagation are decomposed via spin-weighted spherical harmonics. This then allows for groups of coefficients associated with different physical properties to be collected together. The coefficients from operators with no leading-order birefringence are denoted ($c_{F}^{\text{($d$)}}$)$_{njm}^{\text{(0$E$)}}$, where $d$ is the mass dimension, 0 is the spin weight, $E=(-1)^{j}$ is the parity, $n$ is the wavelength dependence, $j$ is the total angular momentum and $m$ is the $z$-component of angular momentum. The camouflage coefficients are then the subset of ($c_{F}^{\text{($d$)}}$)$_{njm}^{\text{(0$E$)}}$ coefficients that are not associated with leading-order vacuum dispersion,
\begin{equation}
(c^{(d)}_{F})^{(0E)}_{njm} = (\voc^{(d)}_{F})^{(0E)}_{njm} - (\voc^{(d)}_{F})^{(0E)}_{(n-2)jm}. \nonumber
\end{equation}
Here, we restrict our attention to $d$=6,8, with $0\leq n\leq$(d-4), $j=n,n-2,\ldots,\geq$0 and $|m|\leq j$. For coefficients with $m\neq 0$ there are real and imaginary components.
\\
The measured fractional beat frequency between the two counter-propagating modes of Fig.~\ref{fig:exp} is
\begin{equation}
\frac{\nu_{\text{beat}}}{\nu}=\sum_{mm'}A_{mm'}e^{im\phi+im'\omega_{\oplus}T_{\oplus}}, \label{eq:nunu}
\end{equation}
where $\phi$ is the angle between the laboratory-frame x axis and geographical south, $\omega_{\oplus}$ is the Earth's sidereal rotation rate and $T_{\oplus}$ is the sidereal time since the last alignment of the experiment with the Sun-Centred Celestial Equatorial Frame (SCCEF), the conventional choice of reference frame for analysis of such experiments within the SME~\cite{Kostelecky:2002}. The $A_{mm'}$ factors contain linear combinations of Lorentz-violating coefficients that a given experiment is sensitive to. They satisfy the relationship $A_{mm'}^{*}=A_{\left(-m\right)\left(-m'\right)}$ and can be calculated from
\begin{equation}
A_{mm'}=\sum_{dnj}\Delta M^{(d)lab}_{njm}d^{(j)}_{mm'}(-\chi)(\voc^{(d)}_{F})_{njm'}.
\label{eq:Amm}
\end{equation}
Here $\Delta M^{(d)lab}_{njm}$ is an experiment-dependent constant that considers the difference between the two counter-propagating modes of Fig.~\ref{fig:exp}, $d^{(j)}_{mm'}$ are the little Wigner matrices and $\chi$ is the co-latitude of the experiment. A detailed guide describing these factors and how to calculate them is provided in section IV of~\cite{mewes1}.
\\
As is standard for the analysis of turntable experiments~\cite{steve1,uwahub} we can model the data as
\begin{align}
\frac{\nu_{\text{beat}}}{\nu}=\sum_{m\geq 0}C_{m}\cos{\left(m\phi\right)}+S_{m}\sin{\left(m\phi\right)}, \label{eq:fit} \\
C_{m}=\sum_{m'\geq 0}C_{mm'}^{C}\cos{\left(m'\omega_{\oplus}T_{\oplus}\right)}+C_{mm'}^{S}\sin{\left(m'\omega_{\oplus}T_{\oplus}\right)}, \label{eq:Cm} \\
S_{m}=\sum_{m'\geq 0}S_{mm'}^{C}\cos{\left(m'\omega_{\oplus}T_{\oplus}\right)}+S_{mm'}^{S}\sin{\left(m'\omega_{\oplus}T_{\oplus}\right)}, \label{eq:Sm}
\end{align}
with the cosine and sine amplitudes of Eqs.~\eqref{eq:Cm} and~\eqref{eq:Sm} given by
\begin{align}
C_{mm'}^{C} = 2\eta_{m}\eta_{m'}\text{Re}(A_{mm'}+A_{m(-m')}), \nonumber \\
C_{mm'}^{S} = -2\eta_{m}\text{Im}(A_{mm'}-A_{m(-m')}), \nonumber \\
S_{mm'}^{C} = -2\eta_{m'}\text{Im}(A_{mm'}+A_{m(-m')}), \label{eq:SC} \\
S_{mm'}^{S} = -2\text{Re}(A_{mm'}-A_{m(-m')}), \label{eq:SS}
\end{align}
with $\eta_{0}=1/2$ and $\eta_{m}=1$ for all other values. Despite the experiment being mounted on a turntable the dataset is effectively stationary with the experiment x-axis permanently aligned parallel to East-West, thereby setting $\phi=\pi/2$. This sets the cosine component of Eq.~\eqref{eq:fit} to zero, so experimental access to Lorentz violating camouflage coefficients is restricted to Eqs.~\eqref{eq:SC} and~\eqref{eq:SS}. In practical terms the beat frequency will be fit to a model containing a linear offset and quadrature components of sidereal harmonics ($m'=1,2,3$),
\begin{align}
\frac{\nu_{\text{beat}}}{\nu}=a+\sum_{m'\geq 0} C_{m'}^{\oplus} \cos{\left(m'\omega_{\oplus}T_{\oplus}\right)} + S_{m'}^{\oplus} \sin{\left(m'\omega_{\oplus}T_{\oplus}\right)}. \label{eq:fitter}
\end{align}
Using equations~\eqref{eq:nunu}-\eqref{eq:fitter} we can explicitly determine the sensitivity of the experiment to higher-order coefficients; results are summarized in Table~\ref{tab:sens}. The amplitudes $C_{1}^{\oplus}$ and $S_{1}^{\oplus}$ provide access to 2 combinations of $d=8$ coefficients that have yet to be bounded.
\begin{table}[t!]
\caption{Sensitivities of sidereal amplitudes to $d=6$ (10$^{-18}$~GeV$^{2}$) and $d=8$ (10$^{-36}$~GeV$^{4}$) camouflage coefficients. Amplitudes are denoted as $C_{m'}^{\oplus}$ for cosine and $S_{m'}^{\oplus}$ for sine; $m'$ is an integer harmonic of $\omega_{\oplus}T_{\oplus}$ as discussed in the main text.}
\centering
\begin{tabular}{c|c}
\hline
Amplitude & Sensitivity \\
\hline
d = 6 & \\
$C_{1}^{\oplus}$ & -0.37 Im($\voc^{6}_{F}$)$_{111}$ \\
$S_{1}^{\oplus}$ & -0.37 Re($\voc^{6}_{F}$)$_{111}$ \\
\hline
d = 8 & \\
$C_{1}^{\oplus}$ & Im($\voc^{8}_{F}$)$_{111}$ - 10.76 Im($\voc^{8}_{F}$)$_{311}$ + 6.00 Im($\voc^{8}_{F}$)$_{331}$ \\
$S_{1}^{\oplus}$ & Re($\voc^{8}_{F}$)$_{111}$ - 10.76 Re($\voc^{8}_{F}$)$_{311}$ + 6.00 Re($\voc^{8}_{F}$)$_{331}$ \\
$C_{2}^{\oplus}$ & 3.25 Im($\voc^{8}_{F}$)$_{332}$ \\
$S_{2}^{\oplus}$ & 3.25 Re($\voc^{8}_{F}$)$_{332}$ \\
$C_{3}^{\oplus}$ & 10.44 Im($\voc^{8}_{F}$)$_{333}$ \\
$S_{3}^{\oplus}$ & 10.44 Re($\voc^{8}_{F}$)$_{333}$ \\
\hline
\end{tabular}
\label{tab:sens}
\end{table}
\section{Results}
\begin{figure}[h]
\centering
\includegraphics[width=0.99\columnwidth]{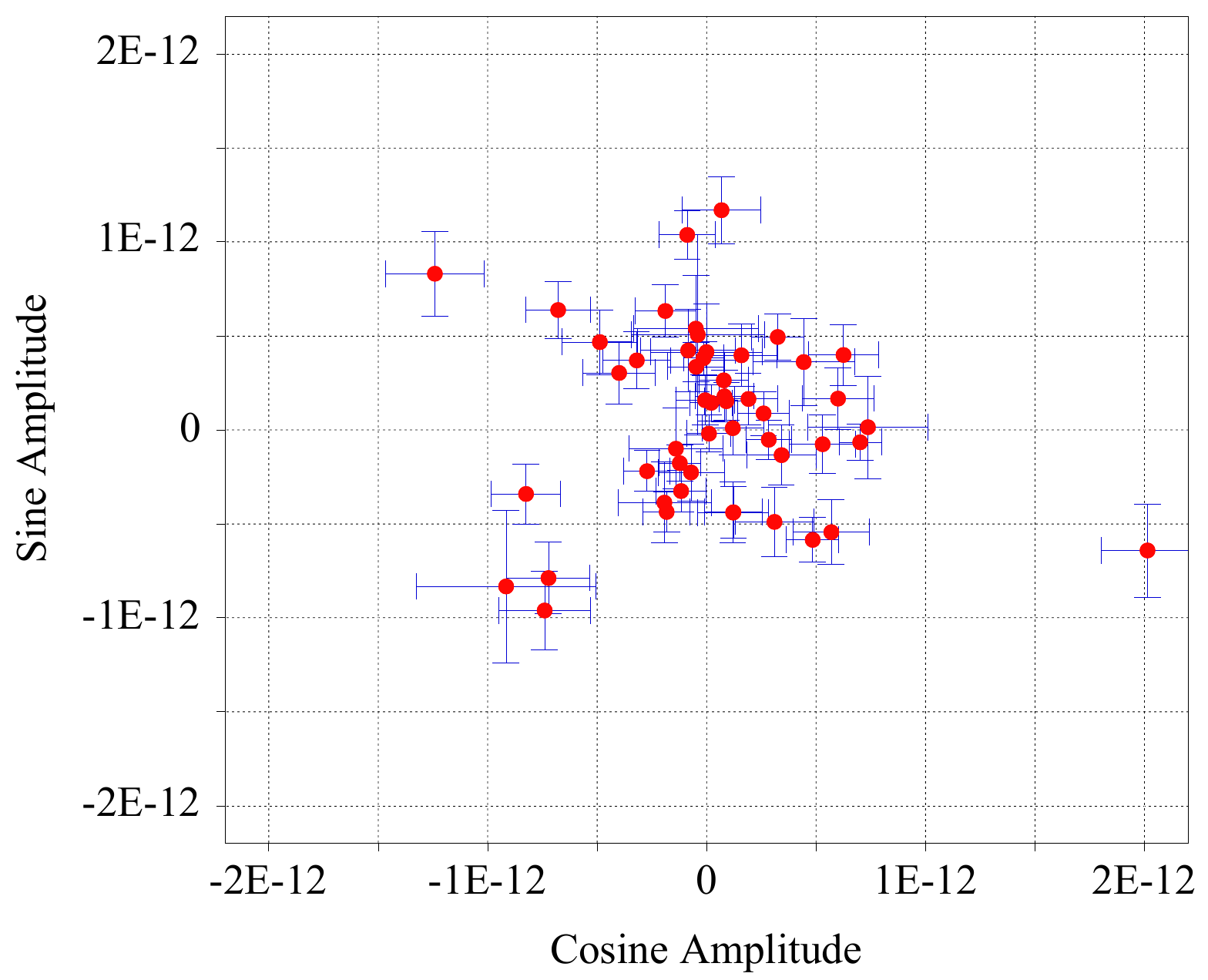}
\caption{Fitted amplitude of fractional frequency shift for the $\omega_{\oplus}$ modulation with 1$\sigma$ error bars. Each data point is obtained by fitting to $\sim$1 day of data, as described in the main text.}
\label{fig:1w}
\end{figure}

\begin{figure}[h]
\centering
\includegraphics[width=0.99\columnwidth]{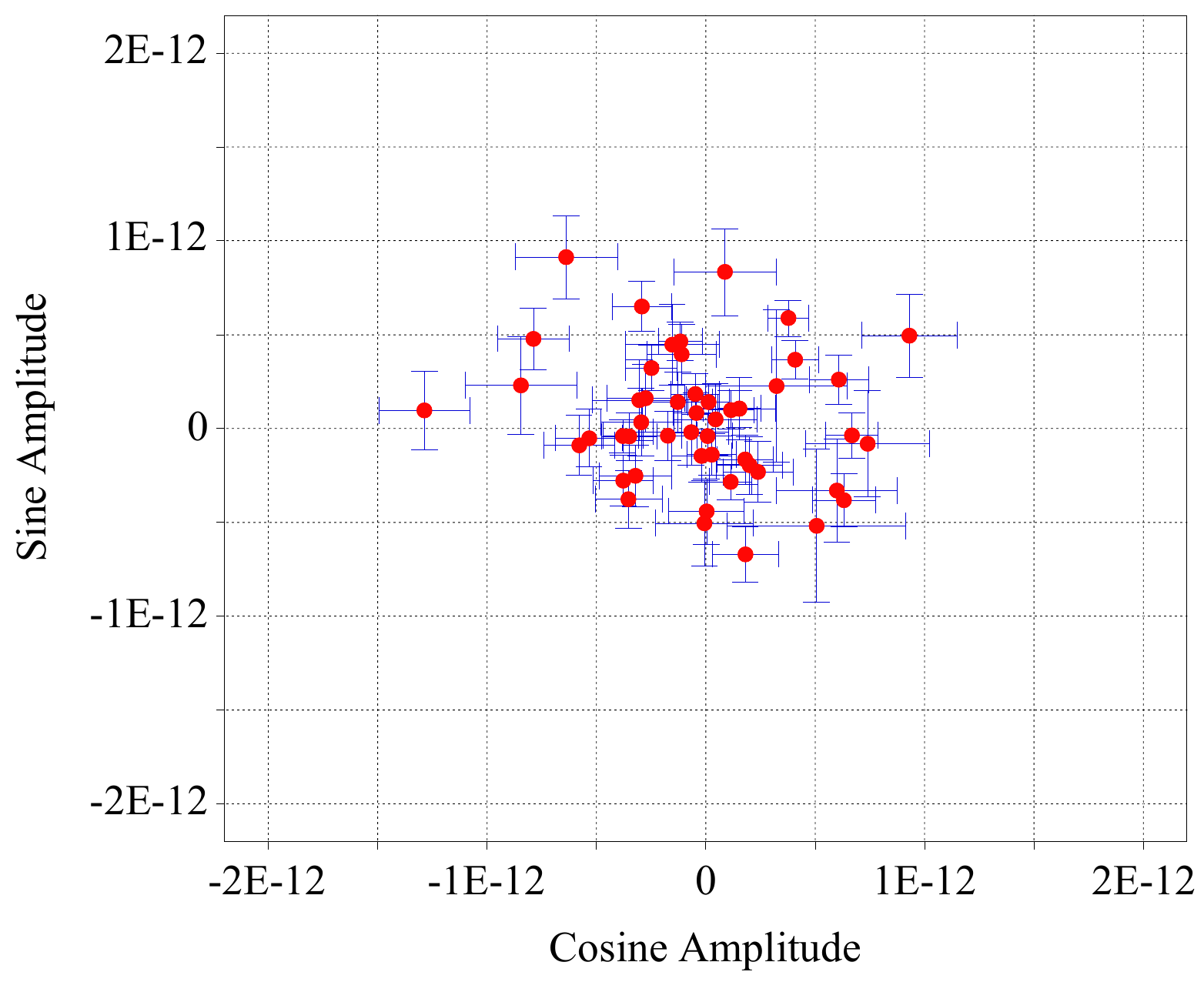}
\caption{Fitted amplitude of fractional frequency shift for the 2$\omega_{\oplus}$ modulation with 1$\sigma$ error bars. Each data point is obtained by fitting to $\sim$1 day of data, as described in the main text.}
\label{fig:2w}
\end{figure}

\begin{figure}[h]
\centering
\includegraphics[width=0.99\columnwidth]{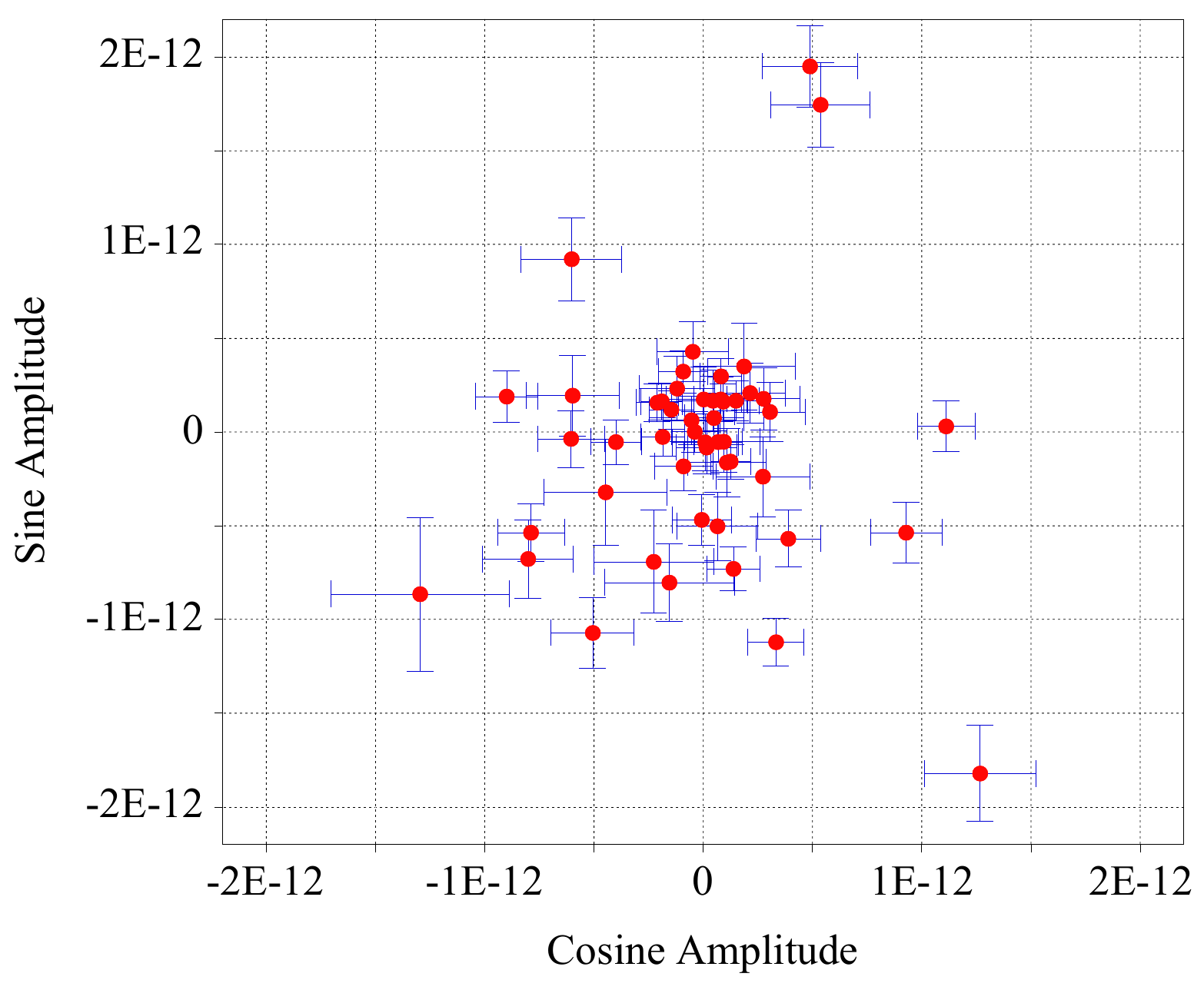}
\caption{Fitted amplitude of fractional frequency shift for the 3$\omega_{\oplus}$ modulation with 1$\sigma$ error bars. Each data point is obtained by fitting to $\sim$1 day of data, as described in the main text.}
\label{fig:3w}
\end{figure}

Data analysis proceeds in a similar manner to previous work~\cite{fred2,uwahub}. The frequency difference between the two counter-propagating modes is averaged while the apparatus is stationary ($\sim$10~minute period). This collection of averaged data points provides us with the $\nu_{beat}$ term in Eq.~\eqref{eq:fit}. Ordinary Least Squares regression is used to simultaneously fit to an offset and quadrature components of modulation of the form $\omega_{\oplus}$, 2$\omega_{\oplus}$ and 3$\omega_{\oplus}$. The fit is performed over subsets of the beat frequency, each comprising 120 points, which is equivalent to one day. This minimizes the standard error of the fit. The magnitude of the error for each frequency of modulation is computed, $\sqrt{C_{m\omega_{\oplus}}^{E}+S_{m\omega_{\oplus}}^{E}}$, and any day of measurement where the magnitude of the error is more than 3$\sigma$ from the mean is discarded. This resulted in the removal of 1 day from the dataset, leaving 50 values for each amplitude which are then combined via an error-weighted average,
\begin{align}
A = \frac{\Sigma\frac{A}{E^2}}{\Sigma\frac{1}{E^2}} \\
E = \sqrt{\frac{1}{\Sigma\frac{1}{E^2}}},
\end{align}
where $A$ is the amplitude and $E$ is the corresponding standard error. Final values with 2$\sigma$ errors are presented in Table~\ref{tab:meas}.
\begin{table}[t!]
\caption{Measured amplitudes of sine and cosine components of $m'\omega_{\oplus}T_{\oplus}$, where $m'$ is denoted in the amplitude coefficient subscript. All values are given $\times$ 10$^{-14}$. Errors are 2$\sigma$.}
\centering
\begin{tabular}{c|c}
\hline
Amplitude & Measurement \\
\hline
$C_{1}^{\oplus}$ & 5.60 $\pm$ 3.96 \\
$S_{1}^{\oplus}$ & 5.73 $\pm$ 3.98 \\
$C_{2}^{\oplus}$ & -1.17 $\pm$ 3.96 \\
$S_{2}^{\oplus}$ & 6.56 $\pm$ 3.98 \\
$C_{3}^{\oplus}$ & -0.44 $\pm$ 3.98 \\
$S_{3}^{\oplus}$ & -2.90 $\pm$ 3.96 \\
\hline
\end{tabular}
\label{tab:meas}
\end{table}
\\
Following the convention established in~\cite{uwahub} we also take the error-weighted average of all $\omega_{\oplus}$ and 2$\omega_{\oplus}$ modulation amplitudes to determine an overall constraint on a shift in the fractional frequency, $\Delta\nu$/$\nu\leq$4.2$\pm$2.0$\times$10$^{-14}$ (95$\%$~C.I.). The purpose of such a value is to provide a single number to represent the fundamental sensitivity of the experiment.
\\
The $S_{2}^{\oplus}$ amplitude is statistically significant at a 3.4$\sigma$ level, however this significance is not constant over the 50 day run of the experiment and only manifests during a $\sim$10 day subset. By plotting the quadrature amplitude fits for each modulation frequency against each other we do not see any persistent signal congruent with a violation of Lorentz invariance (Figs.~\ref{fig:1w},~\ref{fig:2w} and~\ref{fig:3w}).
\\
Now we use Table~\ref{tab:sens} and Table~\ref{tab:meas} to place constraints on higher-order Lorentz-violating coefficients of the SME; results are displayed in Table~\ref{tab:results}. Our constraints for $d=6$ and $d=8$ coefficients already bounded by previous work~\cite{yuta2} are 2 orders of magnitude larger, which is consistent with the differences in absolute sensitivity, duration of measurements and rotation rate between the two experiments. Using our first bounds for two of the linear combinations of $d=8$ coefficients we are able to infer the first constraints for Re($\voc^{8}_{F}$)$_{311}$, Im($\voc^{8}_{F}$)$_{311}$, Re($\voc^{8}_{F}$)$_{111}$ and Im($\voc^{8}_{F}$)$_{111}$. Unfortunately, due to the East-West alignment of the apparatus, the experiment cannot be used to place a limit on the one remaining $d=8$ linear combination, ($\voc^{8}_{F}$)$_{310}$ - 0.093 ($\voc^{8}_{F}$)$_{110}$. However, all of the other parity-odd nonrenormalizable higher-order $d=6$ and $d=8$ coefficients of the photon sector of the SME now have at least one bound constraining them.
\\
Recent work from a parity-even microwave cavity test of Lorentz invariance achieved a fractional frequency sensitivity of 10$^{-18}$~\cite{uwahub}. A planned co-rotating optical and microwave cavity setup with an estimated sensitivity beyond this~\cite{hubcore,nand13} will enable significant improvements to both minimal and parity-even and parity-odd higher-order nonrenormalizable SME coefficient bounds.
\begin{table*}[t!]
\caption{Bounds on higher-order SME camouflage coefficients. $d=6$ coefficients have units GeV$^{-2}$ and $d=8$ coefficients have units GeV$^{-4}$. Comparisons are given to previous work~\cite{yuta2}, combining results from both experiments allows us to infer some individual bounds. Errors are 2$\sigma$.} 
\centering
\begin{tabular}{c|c|c}
\hline
Coefficient & This Work & Previous Work~\cite{yuta2} \\
\hline
d = 6 & & \\
Re($\voc^{6}_{F}$)$_{111}$ & (-1.55 $\pm$ 1.08) $\times$ 10$^{6}$ & (-0.8 $\pm$ 2.2) $\times$ 10$^{3}$ \\
Im($\voc^{6}_{F}$)$_{111}$ & (-1.51 $\pm$ 1.08) $\times$ 10$^{6}$ & (-0.6 $\pm$ 2.2) $\times$ 10$^{3}$  \\
\hline
d = 8 & & \\
($\voc^{8}_{F}$)$_{310}$ - 0.020 ($\voc^{8}_{F}$)$_{110}$ & - &  (-0.2 $\pm$ 3.8) $\times$ 10$^{19}$ \\
Re($\voc^{8}_{F}$)$_{311}$ - 0.020 Re($\voc^{8}_{F}$)$_{111}$ & - & (1.4 $\pm$ 2.6) $\times$ 10$^{19}$ \\
Im($\voc^{8}_{F}$)$_{311}$ - 0.020 Im($\voc^{8}_{F}$)$_{111}$ & - & (0.1 $\pm$ 2.6) $\times$ 10$^{19}$  \\
Re($\voc^{8}_{F}$)$_{311}$ - 0.093 Re($\voc^{8}_{F}$)$_{111}$ & (-5.33 $\pm$ 3.70) $\times$ 10$^{21}$ & - \\
Im($\voc^{8}_{F}$)$_{311}$ - 0.093 Im($\voc^{8}_{F}$)$_{111}$ & (-5.21 $\pm$ 3.68) $\times$ 10$^{21}$ & - \\
Re($\voc^{8}_{F}$)$_{332}$ & (20.19 $\pm$ 12.24) $\times$ 10$^{21}$ & (2.2 $\pm$ 2.6) $\times$ 10$^{19}$ \\
Im($\voc^{8}_{F}$)$_{332}$ & (-3.60 $\pm$ 12.24) $\times$ 10$^{21}$ & (0.2 $\pm$ 2.6) $\times$ 10$^{19}$ \\
Re($\voc^{8}_{F}$)$_{333}$ & (-2.78 $\pm$ 3.80) $\times$ 10$^{21}$ & (-0.1 $\pm$ 3.2) $\times$ 10$^{19}$ \\
Im($\voc^{8}_{F}$)$_{333}$ & (-4.22 $\pm$ 3.82) $\times$ 10$^{21}$ & (-0.1 $\pm$ 3.2) $\times$ 10$^{19}$ \\
\hline
Inferred & & \\
Re($\voc^{8}_{F}$)$_{311}$ & (1.48 $\pm$ 0.98) $\times$ 10$^{21}$ & - \\
Im($\voc^{8}_{F}$)$_{311}$ & (1.43 $\pm$ 0.98) $\times$ 10$^{21}$ & - \\
Re($\voc^{8}_{F}$)$_{111}$ & (7.32 $\pm$ 5.04) $\times$ 10$^{22}$ & - \\
Im($\voc^{8}_{F}$)$_{111}$ & (7.14 $\pm$ 5.00) $\times$ 10$^{22}$ & - \\
\end{tabular}
\label{tab:results}
\end{table*}
\section*{Acknowledgments}
This work was supported by Australian Research Council grant DP130100205.

\section*{References}


\clearpage


\clearpage

\end{document}